\documentclass[a4paper,11pt]{article}
\usepackage{pos}

\title{Lepton flavor physics: some theoretical aspects}

\author*[a,b,c]{Zhi-zhong Xing}

\affiliation[a]{Institute of High Energy Physics, Chinese Academy of Sciences, Beijing 100049, China}
\affiliation[b]{School of Physical Sciences, University 	of Chinese Academy of Sciences, Beijing 100049, China}
\affiliation[c]{Center of High Energy Physics, Peking University, Beijing 100871, China}

\emailAdd{xingzz@ihep.ac.cn}

\abstract{A brief and personal overview of some theoretical aspects of lepton flavor physics
	is presented, with a focus on the canonical seesaw mechanism and Majorana nature of massive
neutrinos.}

\FullConference{12th Neutrino Oscillation Workshop (NOW2024) \\
2-8, September 2024 \\
Otranto, Lecce, Italy \\}

\begin{document}
\maketitle

\section{Historical roles of lepton flavors}

It is well known that the first generation or ``1 G'' of leptons played a crucial role in the development
of Enrico Fermi's effective field theory for beta decays. The most typical example is
$n \to p + e^- + \overline{\nu}^{}_e$, in which both the neutron and the proton consist of the
``1 G'' quarks. Comparing the effective Fermi coupling between the lepton and nucleon
currents with a combination of the two weak-interaction vertices and the $W^-$ mediator of 
$d \to u + e^- + \overline{\nu}^{}_e$ at the quark level, we are left with the ``gauge seesaw''
relation $G^{}_{\rm F} = \sqrt{2} \hspace{0.05cm} g^2/\left(8 M^2_W\right)$. Given the 
numerical inputs $M^{}_W \simeq 80.4
~{\rm GeV}$ and $G^{}_{\rm F} \simeq 1.166 \times 10^{-5} ~{\rm GeV}^{-2}$, one arrives at
$g \simeq 0.65$ for the weak gauge coupling constant, implying that the so-called ``weak'' 
interactions are not really weak at the electroweak scale. A good lesson turns out to be
that a small effective quantity at low energies, such as the Fermi coupling constant 
$G^{}_{\rm F}$, is very likely to originate from some new and heavy degrees of freedom in a
more fundamental theory at much higher energy scales. History repeats itself, as one will
see again and again.

At this point it is worth mentioning that the original form of Fermi's current-current  
interaction is actually $G^{}_{\rm F} \left[\overline{\psi}^{}_p \gamma^{}_\mu \psi^{}_n\right]
\left[\overline{\psi}^{}_e \gamma^\mu \gamma^{}_5 \psi^{}_\nu\right]$~\cite{Fermi:1934hr}, 
a product of the vector current and the axial vector current. Fermi's PhD student Tsung Dao Lee 
once remarked, ``it is curious why Fermi should choose this particular expression, which resembles
the $\rm V$$-$$\rm A$ interaction, but with parity conservation. Unfortunately, by 1956,
when I noticed this, it was too late to ask Fermi.''~\cite{Lee:2001nv}. After parity
violation was discovered in 1957, the $\rm V$$-$$\rm A$ structure of weak interactions was 
established in 1958, paving an important way towards building the standard electroweak theory in 1967.

After the discovery of the muon neutrino in 1962, the two second-generation or ``2 G'' lepton members
went home, making it possible to consider the possibility of lepton flavor mixing~\cite{Maki:1962mu}.
This helped Bruno Pontecorvo to formulate the probability of $\nu^{}_\mu \leftrightarrow \nu^{}_e$
oscillations in 1967. The pursuit of a possible lepton-quark symmetry motivated James Bjorken 
and Sheldon Glashow to propose a new ``charm'' quark with respect to $\nu^{}_\mu$ in 1964, and
its dynamical role became transparent in the Glashow-Iliopolous-Maiani (GIM) mechanism in 1970.
So ``more is (dynamically) different'', just as claimed by Philip Anderson in his principle of
emergence in 1972~\cite{Anderson:1972pca}. The discoveries of the charm quark in 1974 indicated
that a bran new ``GeV'' era began for particle physics, calling for much higher energy machines
to produce much heavier particles. 
 
In 1975 the third and heaviest charged lepton --- $\tau$ was discovered, opening the ``3 G'' era
of leptons and quarks. In this era the Fermilab did the rest to discover the other three 
third-generation or ``3 G'' fermions on behalf of Fermi: the bottom quark in 1977, the top quark in 1995 
and the tau neutrino in 2001. Then the three-family picture of fundamental fermions is complete. 

Similar to the quark sector, the three families of leptons make it possible to discuss leptonic
CP violation in neutrino oscillations either in vacuum~\cite{Cabibbo:1977nk} or in 
matter~\cite{Barger:1980jm}. On the other hand, a global analysis of various neutrino oscillation data
in the standard three-flavor scheme was first made by Gianluigi Fogli and Eligio Lisi's team in 
1994~\cite{Fogli:1993ck} --- an important proof of concept to show its potential (predictive) power,
as can be seen today. 

Going beyond the standard model (SM) in its flavor sector may naturally mean going beyond the 
``3 G'' paradigm of fundamental fermions, especially the ``3 G'' neutrinos, as motivated by
understanding the origin of tiny neutrino masses or by explaining some puzzling anomalies. 
The hypothetical sterile species may be (3 + $n$) with $n = 1, 2, 3, 6, \cdots$. 
However, Steven Weinberg's third law of progress in theoretical physics
should be kept in mind: ``You may use any degrees of freedom you like to describe a physical
system, but if you use the wrong ones, you will be sorry.''~\cite{Guth:1984rq}. In this
sense ``more'' may be stupid. The history of particle physics tells us that {\it a real new 
degree of freedom must be able to help solve at least one fundamental problem and make the 
existing theory more natural, more exact and more powerful}. Otherwise, it would have little
chance to be correct.

\section{Neutrinos are Majorana fermions}

The most natural way to go beyond the SM has implied that massive neutrinos should be the
Majorana fermions. To see this point of view, let us consider the simple structure of the 
standard electroweak theory (without any operators of dimension 5 or higher to assure its
renormalizability) and its economical particle content (without the right-handed neutrino
fields and with only a single Higgs doublet). So neutrinos have to be massless in this case. 
Following the spirit of Weinberg's effective field theory, one may go beyond the SM 
operators of dimension 4 by introducing possible dimension-5 operators in terms of the
available SM fields. It turns out that the dimension-5 operator is unique:
${\cal O}^{}_{\rm W} = \overline{\ell^{}_{\rm L}}\widetilde{H} \widetilde{H}^T \ell^c_{\rm L}/\Lambda$
\cite{Weinberg:1979sa}. After spontaneous electroweak symmetry breaking, such a Weinberg operator 
leads us to small neutrino masses characterized by $\langle H\rangle^2/\Lambda$ with 
$\Lambda \sim 10^{14} ~{\rm GeV}$ being the cut-off scale. 
   
If you believe in the SM and its effective field theory (EFT), you are expected to accept the
Weinberg operator as the most natural and most economical origin of tiny neutrino masses. But
this operator is lepton-number-violating, implying that massive neutrinos must have the 
Majorana nature on the other hand. This observation means that the simplest UV-complete 
realization of ${\cal O}^{}_{\rm W}$ can be done by following three steps: (a)
Adding the right-handed neutrino fields into the SM, but keeping in
mind that they are not the mirror counterparts of the existing left-handed neutrino fields;
(b) allowing the Yukawa interactions between the neutrino fields and the Higgs doublet,
to assure that the neutral and charged fermions have the same right to interact with the
Higgs fields; (c) writing out the self-interaction term of the right-handed neutrino fields 
and their charge-conjugated counterpart, 
namely $\overline{\left(N^{}_{\rm R}\right)^c} M^{}_{\rm R} N^{}_{\rm R}/2$,
a term which respects all the fundamental symmetries of the SM but violates the 
lepton number conservation.  

This canonical seesaw mechanism~\cite{Minkowski:1977sc} is qualitatively elegant
in the sense that it represents a minimal extension of the SM and attributes the smallness 
of active neutrino masses to the largeness of sterile neutrino masses, and thus it
is fully consistent with the spirit of Weinberg's SM effective field theory (SMEFT). 
In this case both light and heavy neutrinos are the Majorana particles contributing on 
an equal footing to the neutrinoless double-beta decays, the most likely place where one 
may see Prof. Majorana in returning home~\cite{Wilczek:2009elb}.  

Let us briefly summarize the pros and cons of the seesaw mechanism. 
(1) Neutrinos have the right to be right-handed to keep a left-right symmetry
--- the most natural and economical extension of the SM with high gains and low costs;
(2) the Majorana nature of massive neutrinos is highly nontrivial as it 
characterizes not only new physics but also new form of matter, in agreement with Weinberg's
EFT prediction; (3) a big bonus is baryogenesis via leptogenesis~\cite{Fukugita:1986hr} 
realized via the CP-violating decays of heavy Majorana neutrinos, making
it possible to {\it kill two birds with one stone}; (4) naturalness of the seesaw mechanism demands 
its scale far above the Fermi electroweak scale, making its direct testability very 
dim \cite{Xing:2009in}; (5) the seesaw-induced fine-tuning issue associated with the Higgs 
mass~\cite{Vissani:1997ys} and the corresponding SM vacuum stability 
issue~\cite{Elias-Miro:2011sqh,Xing:2011aa} need to be taken into account and carefully studied.  
All in all, the canonical seesaw mechanism is most likely to be located in the {\it landscape} 
instead of the {\it swampland} of particle physics~\cite{Vafa:2005ui}.  
 
Recent theoretical progress in studies of the seesaw mechanism includes the complete one-loop 
matching of this mechanism onto the SMEFT~\cite{Zhang:2021jdf}, the complete one-loop 
renormalization-group equations with non-unitarity of the $3 \times 3$ lepton 
flavor mixing matrix in the seesaw EFT framework~\cite{Wang:2023bdw}, and the first complete
and model-independent calculation of the Jarlskog invariant of CP violation for neutrino 
oscillations and the CP-violating asymmetries of heavy Majorana neutrino decays in terms of 
the 18 original seesaw flavor parameters~\cite{Xing:2024xwb}.

\section{Possible lepton flavor symmetries}

The observed quark and lepton flavor mixing patterns tell us that there should have
an approximate up-down parallelism in the quark sector~\cite{Xing:2020ijf}
and an approximate $\mu$-$\tau$ interchange symmetry in the lepton sector~\cite{Xing:2015fdg}.
A lot of flavor symmetries, mainly as an {\it organizing} principle instead of the
{\it guiding} principle, have so far been considered for model building in this regard.
The bottom line for such attempts is that the models should at least be compatible with
current experimental data, as {\it data is King}. Almost all the flavor symmetries under
consideration cannot explain the origin of tiny neutrino masses, so the seesaw
mechanism is usually invoked in building a model of this kind. 

Today it seems that modular invariance is the best seller on the market of flavor 
symmetry model building~\cite{Feruglio:2017spp}. This approach originates from the
string theory framework, in which a single complex modulus $\tau$ is enough to parametrize
the shape of torus and the modular invariant super-potential gives rise to the modular
form of the Yukawa coupling matrices of leptons and quarks as functions of $\tau$. But it
depends on the seesaw mechanism to understand the smallness of active neutrino masses.  
Three immediate questions are in order: (a) What is the physical meaning of the complex 
dimensionless modular parameter $\tau$? (b) How to make the flavor textures of leptons 
and quarks analytically transparent, given the fact that a nonlinear realization of modular 
symmetry causes difficulties in doing analytical approximations? (c) How to naturally 
understand the strong mass hierarchies of charged fermions
in a modular invariant flavor model, given the fact that one has to do a careful 
numerical fitting to convince himself of the validity of the model?      

In contrast, the conventional discrete flavor symmetries can linearly predict the primary
flavor mixing angles with the Clebsch-Gordan coefficients of a given flavor group, and thus
the relevant results are more transparent in physics. But none of the two approaches
is simple and successful. In particular, symmetry breaking is usually more subtle and often
out of control.  

\begin{figure}[t]
	\begin{center}
		\includegraphics[width=0.7\textwidth]{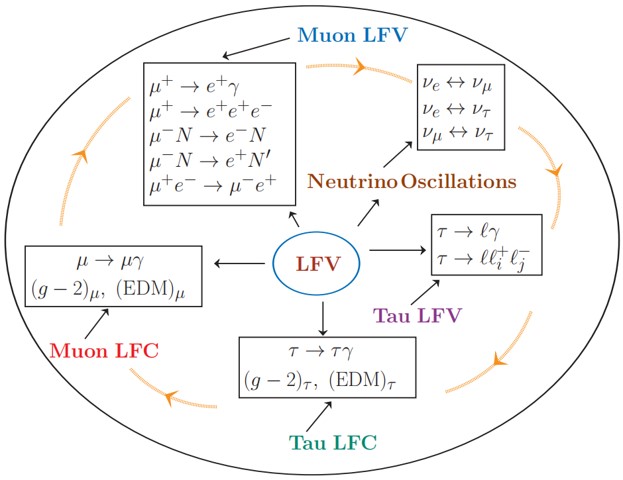}
		\vspace{0cm}
		\caption{Some typical lepton-flavor-violating and lepton-flavor-conserving 
			processes~\cite{IntensityFrontierChargedLeptonWorkingGroup:2013lml}.}
		\label{Fig1}
	\end{center}
	\vspace{-0.4cm}
\end{figure}

Finally, let us emphasize that charged lepton flavors have been playing an important role in 
searching for new physics, as sketched in 
Fig.~\ref{Fig1}~\cite{IntensityFrontierChargedLeptonWorkingGroup:2013lml}. A naive
question is whether the relevant lepton-flavor-violating processes are intrinsically related 
to the mechanisms responsible for the origin of tiny neutrino masses.
The answer is affirmative. In fact, the first full seesaw mechanism was
proposed to calculate the decay rate of $\mu \to e + \gamma$~\cite{Minkowski:1977sc}!
Both light and heavy Majorana neutrinos contribute to such rare processes at the one-loop level, 
and the related lepton flavor mixing matrix elements can therefore be constrained from their 
experimental upper bounds~\cite{Blennow:2023mqx}. 
 
The author is deeply indebted to Eligio Lisi and the organizing committee of NOW2024
for kind invitation and warm hospitality. This work is supported in part by the National
Natural Science Foundation of China under grant No. 12075254.

\end{document}